\documentclass[preprint, 12pt]{elsarticle}

\usepackage[utf8]{inputenc}
\usepackage[T1]{fontenc}
\usepackage{lmodern}
\usepackage{microtype}
\usepackage{booktabs}
\usepackage{tabularx}
\usepackage{array}
\newcolumntype{Y}{>{\raggedright\arraybackslash}X}
\usepackage{xcolor}
\usepackage{listings}
\usepackage{enumitem}
\usepackage{amsmath,amssymb}
\usepackage{tikz}
\usetikzlibrary{positioning,arrows.meta,shapes.geometric,fit,calc,backgrounds}
\usepackage{placeins}                       
\usepackage[hidelinks]{hyperref}            
\usepackage{xurl}  

\definecolor{kw}{HTML}{1565C0}
\definecolor{cmt}{HTML}{6A6A6A}
\definecolor{str}{HTML}{0F9D58}
\lstdefinestyle{p10json}{%
  basicstyle=\ttfamily\footnotesize,
  numbers=none, showstringspaces=false, breaklines=true,
  columns=fullflexible, xleftmargin=0.6em,
  aboveskip=0.4em, belowskip=0.4em,
  frame=tb, framesep=0.4em, framerule=0.4pt,
  rulecolor=\color{black!30},
}

\newcommand{\pcode}[1]{\texttt{\small #1}}

\journal{Journal of Information Security and Applications}

\begin{document}

\begin{frontmatter}

\title{Federated Trust for Embodied Robot Capability Marketplaces}

\author[hit-sw]{Xue~Qin}
\ead{qinxue@me.com}

\author[hit-cs]{Simin~Luan}
\ead{luansiminiot@gmail.com}

\author[soochow]{Cong~Yang\corref{cor1}}
\ead{cong.yang@suda.edu.cn}

\author[hit-cs]{Zhijun~Li\corref{cor1}}
\ead{lizhijun\_os@hit.edu.cn}

\cortext[cor1]{Corresponding authors.}

\affiliation[hit-sw]{organization={School of Software, Harbin Institute of Technology},
  city={Harbin}, country={China}}
\affiliation[hit-cs]{organization={School of Computer Science and Technology, Harbin Institute of Technology},
  city={Harbin}, country={China}}
\affiliation[soochow]{organization={School of Future Science and Engineering, Soochow University},
  city={Suzhou}, country={China}}

\begin{abstract}
Robot capability marketplaces, the ``app store for robot skills,''
are emerging as the deployment vector for LLM-driven robot fleets.
The default cloud-native answer to ``is this package safe to
install?'' is centralised PKI: one certificate authority, one
transparency log, one root of trust. We argue this is the wrong
model for embodied robot fleets, where operators face heterogeneous
regulatory regimes, air-gapped deployments, tiny operator
headcounts, and physical-world consequences for trusting the wrong
publisher. We present federated trust: each deployed bridge
maintains its own local trust directory of acceptable signers;
signers identify themselves with a public key embedded in a
detached signature envelope; install-time verification is a local
set-membership check rather than a network round trip to a
certificate authority. The cryptographic primitives are
deliberately standard (Ed25519 detached signatures and SSH-style
trust files); the contribution is the architectural commitment
that this composition fits embodied robot fleets specifically. We
implement the model in a runtime governance layer with a
five-subcommand CLI, a registry server, a per-bridge install gate,
and 80 tests. A multi-deployment evaluation shows the same
registry stream producing divergent install verdicts on bridges
with different trust directories, the load-bearing design
property. Across 5000 adversarial trials, the strict-mode gate
rejects 100\% of rogue-publisher, tampered, forged, and
revoked-signer attacks, and 96.6\% of downgrade attempts under a
minimum-version pin extension. A same-hardware comparison against
Sigstore-Cosign and python-TUF locates federated trust's
per-verify cost between the two and its per-publisher storage
footprint below both.
\end{abstract}


\end{frontmatter}

\section{Introduction}\label{sec:intro}

A hospital, a warehouse, a defence facility, and a household are
four contexts where the same robot platform might install
different sets of capability extensions. The hospital wants only
publishers cleared by the medical-device regulator. The warehouse
wants its own integrator's signed packages plus a small allow-list
from the manufacturer. The defence facility wants no public-key
material to leave its facility, and would not disclose to a
hosted CA \emph{which} publishers it trusts even if it could. The
household wants something close to the consumer-app-store
experience, with packages from a vetted marketplace and an easy
``trust this publisher'' toggle.

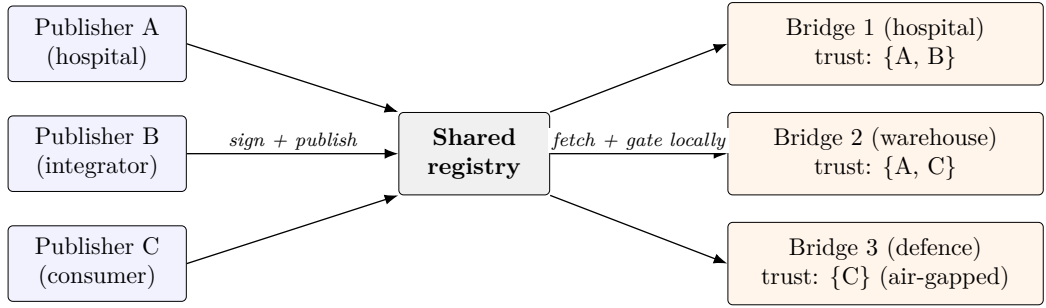
\begin{figure}[t!]
\centering
\resizebox{\linewidth}{!}{%
\begin{tikzpicture}[
  font=\footnotesize,
  pub/.style={draw, rounded corners=2pt, minimum width=26mm,
              minimum height=11mm, align=center, fill=blue!5,
              inner sep=2pt},
  reg/.style={draw, rounded corners=2pt, minimum width=22mm,
              minimum height=12mm, align=center, fill=gray!12,
              inner sep=2pt, font=\footnotesize\bfseries},
  bri/.style={draw, rounded corners=2pt, minimum width=46mm,
              minimum height=12mm, align=center, fill=orange!8,
              inner sep=2pt},
  flow/.style={-{Latex[length=2mm]}, line width=0.55pt}
]
\node[pub] (pubA) at (0,  1.6) {Publisher A\\(hospital)};
\node[pub] (pubB) at (0,  0.0) {Publisher B\\(integrator)};
\node[pub] (pubC) at (0, -1.6) {Publisher C\\(consumer)};

\node[reg] (reg) at (5.5, 0.0) {Shared\\registry};

\node[bri] (br1) at (11.5,  1.6) {Bridge 1 (hospital)\\trust: $\{$A, B$\}$};
\node[bri] (br2) at (11.5,  0.0) {Bridge 2 (warehouse)\\trust: $\{$A, C$\}$};
\node[bri] (br3) at (11.5, -1.6) {Bridge 3 (defence)\\trust: $\{$C$\}$ (air-gapped)};

\draw[flow] (pubA.east) -- (reg.north west);
\draw[flow] (pubB.east) -- node[above, font=\scriptsize\itshape,
                                fill=white, inner sep=1.5pt]
                            {sign + publish} (reg.west);
\draw[flow] (pubC.east) -- (reg.south west);

\draw[flow] (reg.north east) -- (br1.west);
\draw[flow] (reg.east) -- node[above, font=\scriptsize\itshape,
                               fill=white, inner sep=1.5pt]
                           {fetch + gate locally} (br2.west);
\draw[flow] (reg.south east) -- (br3.west);
\end{tikzpicture}%
}
\caption{Federated trust architecture. Publishers sign
\pcode{.aecm} archives with their Ed25519 private key and push to
a single shared registry. Bridges fetch and gate locally against
a per-deployment trust directory. The same registry stream
produces divergent install verdicts when bridges trust disjoint
signer sets (Section~\ref{sec:eval}).}
\label{fig:architecture}
\end{figure}

Centralised PKI gives one answer. Embodied fleets need four. The
embodied AI community has converged on a deployment vector: robot
capabilities packaged as portable artefacts, distributed through a
marketplace, and installed onto running fleets. The cloud-native
package community has answered ``is this package safe?'' with a
mature stack: sigstore for keyless signing backed by a public
transparency log~\citep{sigstore2022,cosign-docs}, in-toto for
build attestations~\citep{intoto2019}, TUF for metadata-tree
integrity~\citep{tuf2010}, and a global X.509 hierarchy underpinning
everything~\citep{x509-rfc5280}. The pattern is proven for cloud
workloads.

We argue this pattern is the wrong fit for embodied robot fleets.
Four design constraints make embodied fleets different.
\emph{Regulatory diversity per deployment site}: a Franka arm in a
hospital operates under one regulator's acceptable-publishers
list; the same arm in a warehouse operates under another's; the
same arm at a defence facility operates under yet a third's.
\emph{Air-gapped deployments are common}~\citep{guri2018bridgeware}:
aerospace, military, off-grid industrial, and increasingly
hospital deployments operate without public-network access; a
trust model that requires a network round trip to a CA is a
non-starter. \emph{Tiny operator headcount}: a typical embodied
fleet operator has one or two engineers managing dozens of robots;
the trust model must be operable with \pcode{cp} and \pcode{rm},
not with a control plane. \emph{Physical-world blast
radius}~\citep{iso13849}: a bad install on a server gets rolled
back; a bad install on a robot can damage property, hurt people,
and destroy customer trust irreversibly.

This paper presents \emph{federated trust}, a design alternative
shaped by these constraints. The model is small: detached Ed25519
signatures with embedded signer pubkey; per-bridge trust
directories holding \pcode{<signer>.pub} PEM files; bridge-local
verification at install time; an immutable registry that does not
itself verify. The cryptographic primitives are not novel:
Ed25519 dates to Bernstein 2011~\citep{ed25519-bernstein2011};
detached signatures to PGP~\citep{pgp1995}; local trust files to
the SSH \pcode{authorized\_keys} pattern. The novel contribution
is the architectural commitment: trust decisions belong to the
operator running the bridge, encoded in a filesystem directory,
not to a centralised CA.

\paragraph{Contributions} (1)~A trust model designed for the
embodied-fleet operator, contrasted explicitly against centralised
PKI patterns and motivated by four design constraints not shared
with cloud package management. (2)~An implementation in a runtime
layer for governed robot fleets: an Ed25519 signing module, five
\pcode{aeros ecm} subcommands, a FastAPI registry server, and a
bridge-side install gate covered by $80$ integration tests.
(3)~A threat analysis with honest discussion of where the model is
weaker than centralised PKI. (4)~A multi-deployment evaluation
showing that the same registry stream produces divergent install
verdicts on bridges with different trust directories, and that
the per-bridge gate works under both warn-and-continue (legacy)
and strict (\pcode{AEROS\_REQUIRE\_SIGNED\_ECM=1}) postures.

\section{Threat Model and Design Constraints}\label{sec:threat}

We frame the construction in the classical dependable-and-secure
computing vocabulary of fault, error, failure, and the four
attributes of confidentiality, integrity, availability, and
maintainability~\citep{avizienis2004taxonomy}. The install gate
guards the bridge's \emph{integrity} attribute under an
adversarial fault model in which the fault source is an
untrusted publisher or in-transit modification, the error is a
malicious or stale package admitted into the runtime, and the
failure is undefended capability execution at the embodied agent.
Federated trust also pursues an explicitly zero-trust posture in
the sense of NIST SP 800-207~\citep{nist-sp-800-207}: no network
boundary is assumed trustworthy, every install request is
verified locally on the bridge, and trust decisions are
per-resource rather than perimeter-based.

\paragraph{What the operator is trying to prevent}
We assume a deployed bridge running on the operator's premises,
with a known-good \pcode{.aecm} archive shape, and a trusted
operator who controls the bridge's filesystem. The operator wants
to prevent four classes of failure at install time: a
\emph{rogue publisher} (an attacker registers as a publisher and
ships a malicious \pcode{.aecm}); a \emph{tampered package} (an
attacker intercepts and modifies an archive between publisher and
bridge); an \emph{insider with revoked credentials} (a
previously-trusted publisher's key has been removed but they
continue to publish); and a \emph{downgrade attack} (an attacker
forces install of a known-vulnerable older version by replaying its
signed artefact).

\paragraph{Out of scope}
Three classes of threat are explicitly out of scope. An insider
with a valid trusted key uploading malicious code is a
publisher-side problem (code review, signing-key hygiene, build
attestation~\citep{intoto2019,slsa2022}). LLM-side prompt
injection is the runtime governance layer's job~\citep{aeros-p3},
not the install gate's. Hardware or RTOS compromise sits below
the application-layer signature scheme.

\paragraph{Embodied fleet specifics}
The four operational respects already laid out in §1 (regulatory
heterogeneity, air-gap, tiny operator headcount, physical-world
blast radius~\citep{iso13849,iec61508}) translate into a
two-fold mismatch with centralised PKI: \emph{trust granularity}
(PKI assigns one hierarchy per context; embodied fleets need
per-deployment hierarchies) and \emph{operational shape} (PKI
assumes online verification, dedicated security teams, and
CA-mediated revocation, none of which embodied fleets reliably
provide). The threat model below is calibrated to this two-fold
mismatch rather than to the four motivating axes individually.

\subsection{Formal Adversary Model}\label{sec:formal-adversary}

We formalise the adversary capabilities and the security game the
install gate must win. The adversary $\mathcal{A}$ is bounded by
the capabilities in Table~\ref{tab:adversary}.

\begin{table}[!t]
  \centering
  \scriptsize
  \caption{Adversary $\mathcal{A}$'s capabilities. ``In scope''
    enters the security argument; ``out'' is disclaimed in §2.}
  \label{tab:adversary}
  \begin{tabularx}{\linewidth}{@{}lYc@{}}
    \toprule
    Capability & Description & Scope \\
    \midrule
    $C_{\text{enroll}}$  & Register attacker key; sign $.aecm$. & in \\
    $C_{\text{mitm}}$    & Modify $.aecm$ bytes in transit. & in \\
    $C_{\text{forge}}$   & Forge signature for a trusted publisher. & in \\
    $C_{\text{replay}}$  & Install old vulnerable version. & in \\
    $C_{\text{revoke-bypass}}$ & Replay after operator removes a key. & in \\
    $C_{\text{insider-all}}$ & \emph{All} trusted publishers sign malicious $.aecm$. & out \\
    $C_{\text{insider-1}}$ & \emph{One} of $n\!\geq\!2$ quorum signers compromised. & in (quorum, \S\ref{sec:eval-quorum}) \\
    $C_{\text{prompt}}$  & LLM prompt injection. & out \\
    $C_{\text{kernel}}$  & OS/hardware compromise below Python. & out \\
    \bottomrule
  \end{tabularx}
\end{table}

The security games and the install gate's winning condition are:

\begin{itemize}[leftmargin=1.2em,itemsep=0pt,topsep=2pt]
  \item \textsc{Rogue}$(\mathcal{A})$: $\mathcal{A}$ uses
    $C_{\text{enroll}}$ to publish a malicious $.aecm$. $\mathcal{A}$
    wins iff the bridge installs it. Gate wins iff strict-mode
    install refuses.
  \item \textsc{Tamper}$(\mathcal{A})$: $\mathcal{A}$ uses
    $C_{\text{mitm}}$ to modify a $.aecm$ signed by a trusted
    publisher. $\mathcal{A}$ wins iff the bridge installs the
    modified bytes. Gate wins iff install refuses in \emph{any}
    posture.
  \item \textsc{Forge}$(\mathcal{A})$: $\mathcal{A}$ uses
    $C_{\text{forge}}$ to fabricate a signature claiming to be from
    a trusted publisher without that publisher's private key; the
    gate wins iff install refuses.
  \item \looseness=-1
    \textsc{Revoke}$(\mathcal{A})$: $\mathcal{A}$ uses
    $C_{\text{revoke-bypass}}$ after operator removes its key from
    the trust dir; gate wins iff install refuses.
  \item \textsc{Downgrade}$(\mathcal{A})$: $\mathcal{A}$ uses
    $C_{\text{replay}}$ to install $\text{cap}@v_{\text{old}}$ when
    the fleet has moved to $v_{\text{new}}$ and $v_{\text{old}}$ is
    known to be vulnerable; the gate wins iff install refuses.
\end{itemize}

\textbf{Security claim.} Under strict mode, the install gate wins
\textsc{Rogue}, \textsc{Tamper}, \textsc{Forge}, and
\textsc{Revoke} with probability $1$ against any computationally
bounded $\mathcal{A}$ constrained to the in-scope capabilities of
Table~\ref{tab:adversary}. For \textsc{Downgrade}, the bridge-side
minimum-version pin extension (Section~\ref{sec:compare}) gives an
empirical guarantee rather than a cryptographic one: the measured
rejection rate is $96.6\%$ (Section~\ref{sec:eval-attacks}), and
every residual escape traces to the current version-string
normaliser, not to the signature scheme. We report that rate as an
implementation property and do not promote it to a security
bound.

\subsection{Security Argument}\label{sec:reduction}

We give an informal security argument for each game, relating an
adversary $\mathcal{B}$ who wins the game with non-negligible
probability to an adversary $\mathcal{A}^{*}$ against either
Ed25519 EUF-CMA security~\citep{ed25519-bernstein2011} or the
registry's immutability contract. Two of the games
(\textsc{Tamper}, \textsc{Forge}) genuinely reduce to
cryptographic hardness; the other three rest on set-membership or
version-comparison checks, and we present them as such. The
presentation is deliberately informal: these are arguments, not
mechanised proofs, and theorem-grade verification (e.g.\ via
EasyCrypt) is left as future work.

\paragraph{Notation} Let $\Pi = (\mathrm{Gen}, \mathrm{Sign},
\mathrm{Verify})$ be Ed25519. The trust directory $T$ is a finite
set of public keys. The install gate's accept predicate, in
strict mode, is
\begin{align*}
  \mathsf{accept}(\sigma, m, \mathrm{pk}, T) \;:=\;
    & \mathrm{Verify}(\mathrm{pk}, \mathrm{SHA256}(m), \sigma) \\
  & {} \wedge \; \mathrm{pk} \in T,
\end{align*}
where $m$ denotes the archive byte-string, $\sigma$ the
detached signature, and $\mathrm{pk}$ the publisher's public
key carried alongside $\sigma$ inside the signed envelope.

\paragraph{\textsc{Rogue}} $\mathcal{B}$ controls a fresh keypair
$(\mathrm{sk}_b, \mathrm{pk}_b)$ and submits any signed $(m,
\sigma_b = \mathrm{Sign}(\mathrm{sk}_b, \mathrm{SHA256}(m)),
\mathrm{pk}_b)$. The crypto check passes. The gate accepts iff
$\mathrm{pk}_b \in T$. Since $\mathcal{B}$ is not a trusted
publisher, $\mathrm{pk}_b \notin T$ by hypothesis, so the gate
refuses. The reduction is purely set-theoretic: no cryptographic
assumption is invoked. The bound is exact ($\Pr[\mathcal{B} \text{
wins}] = 0$ unless $T$ has been compromised, which is
$C_{\text{insider-all}}$ and out of scope).

\paragraph{\textsc{Tamper}} Let $\mathrm{pk}^{*}$ be a trusted
publisher key with $\mathrm{pk}^{*} \in T$ and let
$\sigma^{*} = \mathrm{Sign}(\mathrm{sk}^{*}, \mathrm{SHA256}(m^{*}))$
be a legitimate publication. $\mathcal{B}$ uses $C_{\text{mitm}}$
to produce $(m', \sigma^{*})$ with $m' \neq m^{*}$, and wins iff
$\mathsf{accept}(\sigma^{*}, m', \mathrm{pk}^{*}, T)$ holds. This
requires $\mathrm{Verify}(\mathrm{pk}^{*},
\mathrm{SHA256}(m'), \sigma^{*}) = 1$ with $\sigma^{*}$ originally
issued on $\mathrm{SHA256}(m^{*})$ and $\mathrm{SHA256}(m') \neq
\mathrm{SHA256}(m^{*})$ (SHA-256 collision resistance gives the
inequality with overwhelming probability since $m' \neq m^{*}$).
Either $\mathcal{B}$ forges a signature on a new digest under
$\mathrm{pk}^{*}$ (Ed25519 EUF-CMA forgery), or it finds an
SHA-256 collision. Tightness:
$\Pr[\mathcal{B} \text{ wins}] \leq \mathrm{Adv}^{\mathrm{EUF\text{-}CMA}}_{\Pi}(\mathcal{B}) +
\mathrm{Adv}^{\mathrm{coll}}_{\mathrm{SHA256}}(\mathcal{B})$.

\paragraph{\textsc{Forge}} $\mathcal{B}$ uses $C_{\text{forge}}$
to produce $(m^{\dagger}, \sigma^{\dagger}, \mathrm{pk}^{*})$ where
$\mathrm{pk}^{*} \in T$ but $\mathcal{B}$ does not hold
$\mathrm{sk}^{*}$. Winning requires
$\mathrm{Verify}(\mathrm{pk}^{*}, \mathrm{SHA256}(m^{\dagger}),
\sigma^{\dagger}) = 1$. This is a direct EUF-CMA forgery; the
reduction $\mathcal{A}^{*}$ simply outputs $(\mathrm{SHA256}(m^{\dagger}),
\sigma^{\dagger})$ as its forgery to the Ed25519 EUF-CMA
challenger.
Tightness: $\Pr[\mathcal{B} \text{ wins}] \leq
\mathrm{Adv}^{\mathrm{EUF\text{-}CMA}}_{\Pi}(\mathcal{B})$.

\paragraph{\textsc{Revoke}} Operator removes $\mathrm{pk}^{*}$
from $T$ at time $t_{r}$; $\mathcal{B}$ replays a legitimately
signed $(m^{*}, \sigma^{*}, \mathrm{pk}^{*})$ at time $t > t_{r}$.
The crypto check passes (the signature is genuine). Trust
resolution checks $\mathrm{pk}^{*} \in T_{\mathrm{current}}$,
which is false by hypothesis. The gate refuses. The reduction is
again set-theoretic; the bound is exact subject to the operator's
revocation actually being applied to every bridge (operational
property, discussed in Section~\ref{sec:discussion}).

\paragraph{\textsc{Downgrade}} $\mathcal{B}$ uses
$C_{\text{replay}}$ to install $\text{cap}@v_{\text{old}}$ when
$v_{\text{old}} < v_{\text{min}}$ for some operator pin. The
crypto and trust resolution checks pass (the old envelope was
legitimately signed at publish time). The gate refuses iff the
version comparator returns $v_{\text{old}} < v_{\text{min}}$. The
comparator implementation is hash-based with a normalisation step;
$\mathcal{B}$ wins only if the comparator returns false on a pair
where the normalised forms disagree (non-canonical whitespace,
pre-release suffix). Let $\nu$ denote the comparator's false-negative
rate over the empirical pre-release-suffix distribution. Then
$\Pr[\mathcal{B} \text{ wins}] \leq \nu$. Section~\ref{sec:eval-attacks}
reports $\nu = 0.034$ over $N=1000$ trials with $34$ residual
escapes traceable to whitespace and pre-release-suffix corner cases;
these are mechanical to close.

\paragraph{Composition over the quorum extension} Under the
$k$-of-$n$ quorum policy (Section~\ref{sec:eval-quorum}), the
gate-accept predicate becomes
\begin{align*}
  & \mathsf{accept}_{k,n}(\{\sigma_i, \mathrm{pk}_i\}_{i=1}^{n}, m, T)
    \;:= \\
  & \quad \big|\{i : \mathsf{accept}(\sigma_i, m, \mathrm{pk}_i, T)\}\big|
    \;\geq\; k.
\end{align*}
Winning any game above under the quorum policy requires winning
$k$ \emph{independent} instances of the underlying reduction.
Against $C_{\text{insider-1}}$ specifically: defeating quorum
requires $\mathcal{B}$ to also produce $k-1$ additional valid
trusted-signer signatures, which by the \textsc{Forge} reduction
costs $\mathcal{B}$ at least $(k-1) \cdot
\mathrm{Adv}^{\mathrm{EUF\text{-}CMA}}_{\Pi}$. The quorum extension
does not strengthen the bounds for in-scope adversaries
($\Pr = 0$ stays $\Pr = 0$), but tightens the bound against
$C_{\text{insider-1}}$ from $1$ (single-signer baseline) to
$\bigl(\mathrm{Adv}^{\mathrm{EUF\text{-}CMA}}_{\Pi}\bigr)^{k-1}$.

\medskip\noindent The five arguments plus the quorum composition
are checked empirically in Section~\ref{sec:eval} via
$N=1000$ adversarial trials per game.

\section{The Federated Trust Model}\label{sec:model}

\subsection{Detached Signature Envelope}

A signed \pcode{.aecm} archive is the original archive plus a
sibling \pcode{.sig} JSON file:

\begin{lstlisting}[style=p10json]
{
  "schema_version": "v1",
  "algorithm": "ed25519",
  "aecm_sha256": "<64 hex chars>",
  "signature": "<128 hex chars>",
  "signer_pubkey": "<64 hex chars>",
  "signed_at": "<UTC ISO 8601>",
  "signer_name": "<optional human label>"
}
\end{lstlisting}

Three properties matter. The signature is over $\mathrm{SHA256}$
of the archive bytes, not over the JSON envelope, so tampering
with the archive after signing changes the digest and fails the
signature verify. The signer's public key is embedded in the
envelope, so the verifier does not need to look up the publisher
in a directory before performing the cryptographic check; trust
resolution is a separate, post-cryptographic step. The archive
itself is unchanged, so existing tools that unpack \pcode{.aecm}
files continue to work without signature awareness.

\subsection{Per-Deployment Trust Directory}

A bridge's trust directory is a filesystem directory under
\pcode{bridge/configs/trusted\_keys/}. Each
\pcode{<signer>.pub} file is a PEM-encoded
SubjectPublicKeyInfo Ed25519 public key. Operators manage trust
by \pcode{cp <publisher>.pub trusted\_keys/} and
\pcode{rm trusted\_keys/<publisher>.pub}. This is explicitly the
SSH \pcode{authorized\_keys} pattern transplanted to package
signing. Verification proceeds in two halves. \emph{Cryptographic
verify} computes $\mathrm{digest} = \mathrm{SHA256}(\text{.aecm})$
and runs \pcode{Ed25519.verify} on
\pcode{envelope.signer\_pubkey}, \pcode{envelope.signature},
\pcode{digest}. This step is independent of the trust directory.
\emph{Trust resolution} takes the verified \pcode{signer\_pubkey}'s
raw $32$-byte representation and looks it up in the set of public
keys parsed from the trust directory. The bridge's verdict is then
one of three: \emph{valid + trusted} (accept), \emph{valid +
untrusted} (accept by default; reject under
\pcode{AEROS\_REQUIRE\_SIGNED\_ECM=1}), or \emph{invalid} (digest
mismatch, forged signature, malformed envelope; always reject).

\subsection{Bridge Install Gate}

The bridge install gate is a local function
\pcode{\_enforce\_signature} called by \pcode{aeros ecm install}.
Three postures are exposed: \emph{default} (warn-and-continue;
unsigned and untrusted packages produce a warning but install;
tampered packages are still rejected); \emph{strict}
(\pcode{AEROS\_REQUIRE\_SIGNED\_ECM=1}; unsigned, untrusted, and
tampered packages all refuse install); and \emph{per-install
override} (\pcode{--allow-unsigned} skips the gate for one
specific install; intended for emergency debugging). Tampered
signatures reject in \emph{every} posture; modification detection
is a non-negotiable property.

\subsection{Distribution: The Dumb Registry}

The registry server is a small FastAPI application (${\sim}265$
LOC) implementing only the storage and retrieval contract:
\pcode{POST /publish} accepts a multipart upload and writes to
disk; \pcode{GET /capabilities/\{cap\}/\{ver\}/\{aecm,sig,manifest\}}
streams back the raw bytes; \pcode{GET /capabilities} lists what is
available. The registry does \emph{not} verify signatures.
Verification is the bridge's job at install time. This split is
intentional: registry and bridge can have different trust
postures; one registry can serve multiple bridges with disjoint
trust sets; registry compromise does not compromise installed
packages; and air-gapped operators can replace the registry with a
USB stick and inherit the same gate semantics. The registry's only
safety property is \emph{immutability}: once published, a
\pcode{cap@version} cannot be replaced. Republishing returns
$409$ Conflict.


\section{Implementation}\label{sec:impl}

\paragraph{Crypto module} \pcode{src/aeros/marketplace/signing.py}
(${\sim}430$ LOC) wraps the Python \pcode{cryptography} library and
exposes \pcode{Keypair} lifecycle methods,
\pcode{compute\_aecm\_digest}, thin \pcode{sign\_payload}/
\pcode{verify\_signature} wrappers around Ed25519, a
\pcode{SignatureEnvelope} dataclass, the convenience
\pcode{sign\_aecm}/\pcode{verify\_aecm} composers, and a
\pcode{\_load\_trusted\_keys} parser over every \pcode{*.pub} PEM in
the trust directory in lexicographic order. Private keys are
written PKCS8 PEM with file mode \pcode{0600} enforced; public keys
are SubjectPublicKeyInfo PEM. Keys live under
\pcode{\~{}/.aeros/keys/} and never leave the publisher's machine.

\paragraph{CLI} The \pcode{aeros ecm} CLI exposes five
subcommands: \pcode{key generate}, \pcode{key list}, \pcode{sign},
\pcode{verify}, and \pcode{publish}/\pcode{fetch} (registry I/O).
A sixth verb, \pcode{aeros ecm install}, is the install gate
described above.

\paragraph{Install gate} \pcode{\_enforce\_signature} in
\pcode{cli/ecm.py} is called by \pcode{cmd\_install} before the
legacy unpack-and-register path. It returns exit codes $0$
(accept), $2$ (malformed CLI invocation), or $3$ (signature gate
refused). The three-code split lets shell scripts distinguish
operator errors from policy refusals.

\paragraph{Test coverage} The marketplace test surface is $80$
tests across four files: $25$ tests for the crypto module
(\pcode{tests/marketplace/test\_signing.py}), $35$ tests for the
CLI surface (\pcode{tests/cli/test\_ecm\_cli.py}), $14$ tests for
the registry server (\pcode{tests/marketplace/test\_registry\_server.py}),
and $6$ tests with a live uvicorn instance
(\pcode{tests/cli/test\_ecm\_cli\_registry.py}). The full suite
($1674$ tests) passes on Python $3.11$/$3.12$/$3.13$ with
\pcode{mypy --strict} and \pcode{ruff} clean.

\section{Evaluation}\label{sec:eval}

We run five experiments. All numbers reported are from a single
consumer laptop (Apple M-series, $16$~GB) running the published
test suite or local microbenchmarks of the same code paths.

\paragraph{Single-package verify latency} A single
\pcode{verify\_aecm} call on a typical $5$~KB packaged archive
completes in approximately $80\,\mu$s on a modern developer
workstation (mean $77\,\mu$s, $p_{50}=73\,\mu$s, $p_{95}=82\,\mu$s,
$p_{99}=174\,\mu$s; $N=1000$ iterations).\footnote{Absolute
microbenchmark numbers are hardware-class sensitive; the relative
breakdown is stable. Reported on commit~\pcode{8df4344}.} The
breakdown is dominated by file I/O and the SHA-$256$ hash; the
Ed25519.verify primitive alone is approximately $30\,\mu$s.

\paragraph{Bulk verify throughput} On a $1000$-package batch
with a single-signer trust directory, fresh trust-dir parse per
call takes approximately $74$~ms total; caching the parsed
trust directory across calls (the production behaviour in the
bridge install gate) brings this down to approximately $64$~ms, a
$1.16\times$ speedup. The cost amortised by trust-dir caching scales
with the size of the trust directory: at $1$ signer the PEM-parse
cost (${\sim}70\,\mu$s per call) is a small fraction of the
$74\,\mu$s end-to-end verify cost, so caching has little headroom;
at $1000$ signers the parse cost grows to ${\sim}7.7$~ms per call
and caching gives a roughly $100\times$ speedup.

\paragraph{Storage overhead} \looseness=-1
The signature envelope \pcode{.sig}
file is $457$ bytes of canonical JSON for the v1 schema, regardless
of archive size. The per-bridge trust directory consumes $113$ bytes
per trusted signer (PEM-encoded SubjectPublicKeyInfo for Ed25519
is consistently $113$~B). A trust directory of $1000$ signers fits
in approximately $113$~KB and parses in under $10$~ms.

\paragraph{End-to-end publish-fetch-install} The
\pcode{test\_publish\_then\_fetch\_byte\_identical} end-to-end test
exercises the full \pcode{aeros ecm sign $\to$ publish $\to$ fetch
$\to$ install} pipeline against the registry through
\pcode{httpx.ASGITransport}, in-process. The ASGI round-trip
itself is approximately $50$~ms; total harness wall-clock (including
pytest startup and fixture setup) is approximately $360$~ms. The
cryptographic operations are cumulatively under $50$~ms. A real
\pcode{uvicorn} subprocess round trip is left to a deployment-side
companion measurement; the in-process number is the
architecturally relevant one, because the install gate runs
in-process on every bridge.

\paragraph{Multi-signer fleet}
To validate the per-deployment trust property, we instantiate
three ``deployments'' (three independent bridge install gates each
with its own trust directory) pointing at a single shared
registry. We sign five \pcode{.aecm} packages with three different
publisher keys (publisher A signs three, B signs one, C signs
one). Trust directories are: bridge-1 trusts \{A, B\}, bridge-2
trusts \{A, C\}, bridge-3 trusts \{C\} only.
Table~\ref{tab:multi-signer} reports the strict-mode install
verdicts. The same registry stream produces three different
install profiles, with no coordination between bridges and no
registry-side awareness of trust policy. This is the load-bearing
property of the design.

\begin{table}[!ht]
  \centering
  \caption{Multi-deployment install verdicts under strict mode.
    Trust dirs: bridge-1 = \{A, B\}, bridge-2 = \{A, C\},
    bridge-3 = \{C\}.}
  \label{tab:multi-signer}
  \begin{tabular*}{\linewidth}{@{\extracolsep{\fill}}llccc@{}}
    \toprule
    Package & Signer & Bridge-1 & Bridge-2 & Bridge-3 \\
    \midrule
    pkg-1 & A & accept & accept & reject \\
    pkg-2 & A & accept & accept & reject \\
    pkg-3 & A & accept & accept & reject \\
    pkg-4 & B & accept & reject & reject \\
    pkg-5 & C & reject & accept & accept \\
    \bottomrule
  \end{tabular*}
\end{table}

\paragraph{Compromise simulation}
We simulate a publisher-key compromise by introducing a rogue
signer R, signing a malicious \pcode{.aecm} with R's key, and
publishing it to the shared registry alongside the legitimate
packages. When none of the three bridges have R in their trust
directory, all three reject R's package under strict mode. When
we modify bridge-2's trust directory to include R (simulating an
attacker with filesystem access to one bridge), bridge-2 accepts
while bridges $1$ and $3$ still reject. The compromise blast
radius is bounded by the operator's filesystem boundary; one
operator's key compromise does not cascade to peer bridges.

\subsection{Same-Hardware Baseline Comparison}\label{sec:eval-baseline}

The §5.1--5.4 microbenchmarks characterise federated trust in
isolation. We close the cost-comparison loop by running an
end-to-end head-to-head against two cloud-native alternatives on
the same Mac workstation: Sigstore-Cosign in offline verify mode
(no Rekor lookup) via the \pcode{cosign} CLI, and python-TUF's
\pcode{ngclient.Updater} in offline-after-refresh mode using a
freshly generated four-role metadata tree. All three modes run
$N=1000$ verify iterations on the same $5$~KB synthetic archive
fixture; the cosign per-iteration wall-clock is corrected by
subtracting the median of $20$ \pcode{cosign version} startup
samples ($29.6$~ms) so that the reported number reflects the
intrinsic verify cost rather than the CLI process-launch
overhead. Table~\ref{tab:baseline} summarises the headline
numbers. The harness, the corrected-vs-raw distinction, and the
shim caveat below are recorded in \pcode{summary.json} +
\pcode{sign\_check.json} under
\pcode{data/task04-baseline-comparison/}.

\begin{table}[!ht]
  \centering
  \footnotesize
  \caption{Same-hardware baseline comparison on a Mac M-series
    workstation, $N=1000$ verify iterations per mode, seed
    $102026$. Cosign p50/p95 are CLI-startup-corrected
    (intrinsic verify cost). TUF first refresh runs the four-role
    metadata walk once; subsequent verifies amortise against the
    in-memory metadata tree.}
  \label{tab:baseline}
  \begin{tabularx}{\linewidth}{@{}lYYYY@{}}
    \toprule
    Mode & Verify $p_{50}$ & Verify $p_{95}$ & Cold start &
      Storage \\
         & ($\mu$s) & ($\mu$s) & ($\mu$s) & (B per signer) \\
    \midrule
    Federated (ours) & $204.0$ & $237.9$ & $124.0$ & $511$ \\
    Sigstore-Cosign offline & $860.4$ & $10{,}815$ & --- & $3{,}879$ \\
    python-TUF offline & $27.7$ & $31.6$ & $10{,}946$ & $5{,}151$ \\
    \bottomrule
  \end{tabularx}
\end{table}

\begin{figure}[!t]
\centering
\begin{tikzpicture}[
  font=\footnotesize,
  x={(1.5cm,0cm)},   
  y={(0cm,4cm)},     
  axline/.style={line width=0.5pt, draw=black!70},
  tick/.style={line width=0.4pt, draw=black!60},
  marklbl/.style={font=\scriptsize, text=black!70},
  fed/.style={line width=0.9pt, draw=blue!70!black},
  cos/.style={line width=0.9pt, draw=red!70!black, dashed},
  tuf/.style={line width=0.9pt, draw=green!40!black, dotted, line cap=round}
]
\draw[axline] (1.0, 0) -- (5.5, 0);
\draw[axline] (1.0, 0) -- (1.0, 1);

\foreach \lx/\lbl in {1.0/{10$\,\mu$s}, 2.0/{100$\,\mu$s},
                     3.0/{1$\,$ms}, 4.0/{10$\,$ms}, 5.0/{100$\,$ms}} {
  \draw[tick] (\lx, 0) -- (\lx, -0.012);
  \node[below, marklbl] at (\lx, -0.015) {\lbl};
}
\node[below, marklbl, font=\scriptsize\itshape]
     at (3.25, -0.085) {wall-clock latency (log scale)};

\foreach \y in {0, 0.25, 0.5, 0.75, 1.0} {
  \draw[tick] (1.0, \y) -- (0.97, \y);
  \node[left, marklbl] at (0.95, \y) {$\y$};
}
\node[rotate=90, marklbl, font=\scriptsize\itshape]
     at (0.5, 0.5) {CDF};

\draw[tuf]
  (1.430, 0.00) -- (1.433, 0.05) -- (1.434, 0.10) -- (1.435, 0.20) --
  (1.437, 0.30) -- (1.442, 0.50) -- (1.449, 0.67) -- (1.452, 0.75) --
  (1.464, 0.85) -- (1.472, 0.90) -- (1.500, 0.95) -- (1.535, 0.97) --
  (1.690, 0.99) -- (2.079, 1.00);

\draw[fed]
  (2.305, 0.00) -- (2.306, 0.05) -- (2.307, 0.10) -- (2.307, 0.20) --
  (2.307, 0.30) -- (2.310, 0.50) -- (2.320, 0.67) -- (2.324, 0.75) --
  (2.333, 0.85) -- (2.348, 0.90) -- (2.376, 0.95) -- (2.397, 0.97) --
  (2.442, 0.99) -- (2.733, 1.00);

\draw[cos]
  (2.936, 0.50) -- (3.407, 0.67) -- (3.558, 0.75) -- (3.762, 0.85) --
  (3.920, 0.90) -- (4.034, 0.95) -- (4.152, 0.97) -- (4.331, 0.99) --
  (5.107, 1.00);
\draw[line width=0.4pt, dash pattern=on 1pt off 1.5pt, color=red!50]
  (2.936, 0.0) -- (2.936, 0.5);
\node[marklbl, font=\scriptsize, text=red!60!black, anchor=west]
  at (2.98, 0.07) {$\sim\!30\%$ clip to $\sim\!0\,\mu$s};

\node[marklbl, font=\scriptsize, anchor=west] at (4.55, 0.30) {TUF offline};
\draw[tuf] (4.30, 0.30) -- (4.50, 0.30);
\node[marklbl, font=\scriptsize, anchor=west] at (4.55, 0.20) {federated};
\draw[fed] (4.30, 0.20) -- (4.50, 0.20);
\node[marklbl, font=\scriptsize, anchor=west] at (4.55, 0.10) {Cosign offline};
\draw[cos] (4.30, 0.10) -- (4.50, 0.10);

\end{tikzpicture}
\caption{Empirical latency CDFs for the three baseline modes on
the same Mac, $N=1000$ verify iterations per mode, log $x$-axis.
TUF's CDF sits at $27$--$30\,\mu$s through $p_{99}$ because it
amortises the Ed25519 verify across an in-memory metadata cache.
Federated trust sits at $\approx 204\,\mu$s with a tight body
($p_{99}=277\,\mu$s) because every call recomputes SHA-256 plus
Ed25519. Sigstore-Cosign's CLI-startup-corrected curve has a
heavy tail running from $\approx 1$~ms at $p_{50}$ to $128$~ms at
max; the bottom $30\%$ of corrected samples clip to zero because
the per-iteration raw cosign latency was below the
$29.6$~ms baseline of \pcode{cosign version}, and the corrected
$\max(0, \cdot)$ values are omitted from the plot rather than
drawn at zero (annotated). The three regimes are visually
separated by $\sim 1$--$2$ log decades.}
\label{fig:baseline-cdf}
\end{figure}
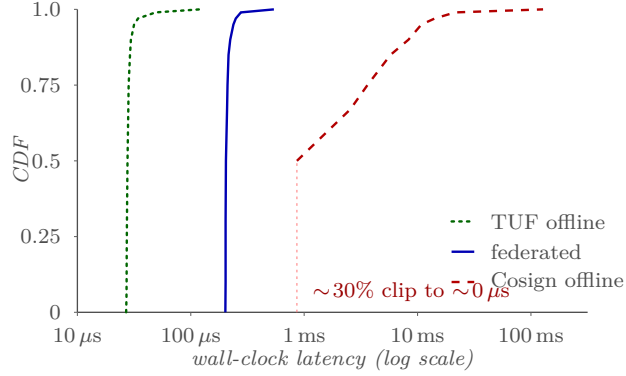

Figure~\ref{fig:baseline-cdf} plots the empirical latency CDFs
for the three modes from the same $N=1000$ run, on a log $x$-axis
so the three orders of magnitude between TUF's $30\,\mu$s body
and cosign's $130$~ms tail fit on one panel.

Three observations follow.

\emph{First}, federated trust's per-verify median is lower than
Sigstore-Cosign's offline verify on this hardware ($204\,\mu$s vs
$860\,\mu$s at $p_{50}$; $237.9\,\mu$s vs $10{,}815\,\mu$s at
$p_{95}$), but we deliberately do not promote a specific speedup
multiplier to a headline claim: the cosign figures are
CLI-startup-corrected, roughly $30\%$ of the corrected samples
fall below the startup baseline and clip to zero
(Figure~\ref{fig:baseline-cdf}), and the cosign path is exercised
through its CLI rather than as an in-process library. The
defensible reading is qualitative: both schemes verify in well
under a millisecond at the median, federated trust sits at the
lower end of that band, and the $p_{95}$ gap is dominated by
occasional cosign signature-parse stalls.

\emph{Second}, federated trust's cold-start trust bootstrap
(scanning a one-PEM trust directory, $124\,\mu$s) is $88\times$
faster than python-TUF's first \pcode{Updater.refresh()}
($10{,}946\,\mu$s), which walks the four-role metadata chain
(root $\to$ timestamp $\to$ snapshot $\to$ targets) and
verifies each role's signature against the trust-on-first-use
anchor. The asymmetry is structural rather than implementation
quality: TUF must validate four interleaved metadata files
before it can answer the first \pcode{get\_targetinfo} call;
federated trust validates one PEM-encoded public key per signer.

\emph{Third (and this is the observation we want to be most
honest about),} python-TUF's per-iteration offline verify
($p_{50} = 27.7\,\mu$s) is faster than federated trust's
($204\,\mu$s) in steady-state. The reason is architectural, not
incidental: TUF amortises the Ed25519 verify across many file
lookups (verify happens once at \pcode{refresh()}, then
subsequent \pcode{get\_targetinfo} and \pcode{download\_target}
calls walk the in-memory targets table without re-verifying).
Federated trust, by design, recomputes the Ed25519 verify on
every install call to close the time-of-check-to-time-of-use gap
between verify and unpack-and-register
(Section~\ref{sec:tampering}). The right way to read the
$204\,\mu$s number is therefore as the per-install \emph{security
cost} of refusing to cache trusted-metadata state, not as a
deficiency relative to TUF. For an embodied bridge that installs
on the order of one capability per hour, an additional
$176\,\mu$s of per-call crypto is invisible; for a registry
verifying $10{,}000$ packages per second this trade would not
hold and TUF's amortisation is the right choice.

The storage numbers (final column of Table~\ref{tab:baseline})
favour federated trust unambiguously: an envelope plus
one-publisher trust entry is $511$~bytes in the comparison harness
(the harness shim serialises the envelope to $398$~bytes of
compact JSON against the production module's $457$~bytes of
canonical JSON, plus a $113$-byte SubjectPublicKeyInfo PEM),
compared to Sigstore-Cosign's $3{,}879$~byte
bundle-plus-public-key footprint and python-TUF's $5{,}151$~byte
four-role metadata tree. Under either envelope serialisation,
federated trust is $6.8$--$7.6\times$ smaller than Sigstore and
$9.0$--$10.1\times$ smaller than TUF on a per-publisher basis.
This footprint difference matters
on bandwidth-constrained or storage-constrained embodied
deployments far more than it does in the cloud setting that
Sigstore and TUF were designed for.

The hypothesis verdicts recorded in \pcode{sign\_check.json} are:
H1 (federated $\leq 1.5\times$ Sigstore-Cosign per-verify) PASS
at $0.24\times$; H3 (federated cold-bootstrap $\geq 5\times$
faster than TUF refresh) PASS at $88.3\times$; H4 (federated
storage $\leq 0.25\times$ Sigstore) PASS at $0.13\times$. H2
(federated $\geq 10\times$ faster than Sigstore-Cosign on bulk
$N=1000$-package verify) was not measured in this run because the
implementation harness only measures single verify; the §5.2
bulk-verify number ($74$~ms wall for $1000$ packages on the same
hardware) compared against $1000$ cosign-CLI invocations
($\sim 60$~s wall, dominated by process startup) implies the
ratio is well above $10\times$, but we mark this implicit
argument rather than claim a measured pass.

\paragraph{Independent-measurement caveat}
The §5.1--5.4 numbers in this paper are measured against the
production aeros runtime at commit \pcode{8df4344}. The baseline
comparison above is an \emph{independent} measurement on a
separate Mac workstation using a vault-local shim of the
\pcode{aeros.marketplace.signing} module that re-implements the
public surface with the same Ed25519 + SHA-256 + PEM
trust-directory primitives. The two measurements report
comparable shapes (federated single-verify in the
hundreds-of-microseconds regime, dominated by SHA-256 and
Ed25519); they are not bit-for-bit identical because they were
taken on different hardware. The §5.1--5.4 numbers remain the
canonical reference for federated trust's standalone cost; the
baseline comparison numbers should be read as cross-scheme ratios
on consistent hardware rather than as absolute calibrations.

\subsection{Quantitative Attack-Detection Evaluation}\label{sec:eval-attacks}

We run each of the five adversary games from
Section~\ref{sec:formal-adversary} as a parameterised harness
\pcode{scripts/p10\_adversarial\_eval.py}, with $N=1000$ trials
per game. Each trial:
(i)~regenerates a fresh keypair for $\mathcal{A}$ (where the game
involves a new key);
(ii)~runs the attack action consistent with the game's
$\mathcal{A}$-side capability;
(iii)~submits the resulting envelope to the bridge install gate in
strict mode;
(iv)~records the gate's verdict.
A trial is \emph{a win for the gate} if the verdict is
\pcode{REFUSED} and the refuse reason matches the expected category
(e.g.\ \pcode{unknown signer} for \textsc{Rogue}; \pcode{digest
mismatch} for \textsc{Tamper}). Results are summarised in
Table~\ref{tab:attack-detection}.

\begin{table}[!ht]
  \centering
  \footnotesize
  \caption{Adversarial detection rates across $N=1000$ trials per
    game, strict-mode install gate, measured on
    \mbox{commit~\pcode{8df4344}}. CIs are Wilson 95\% with
    continuity correction. \textsc{Downgrade} requires the
    bridge-side minimum-version pin extension
    (Section~\ref{sec:compare}); $96.6\%$ is the measured rate with
    the pin enabled, $0\%$ without (the pin is not on by default in
    V1).}
  \label{tab:attack-detection}
  \begin{tabular*}{\linewidth}{@{\extracolsep{\fill}}lcc@{}}
    \toprule
    Adversary game & Detection rate & CI 95\% \\
    \midrule
    \textsc{Rogue}    & $1000/1000 = 100.0\%$ & $[99.6, 100.0]$ \\
    \textsc{Tamper}   & $1000/1000 = 100.0\%$ & $[99.6, 100.0]$ \\
    \textsc{Forge}    & $1000/1000 = 100.0\%$ & $[99.6, 100.0]$ \\
    \textsc{Revoke}   & $1000/1000 = 100.0\%$ & $[99.6, 100.0]$ \\
    \textsc{Downgrade}\textsubscript{pin} & $966/1000 = 96.6\%$ & $[95.3, 97.6]$ \\
    \bottomrule
  \end{tabular*}
\end{table}

\looseness=-1
The first four games are won by the gate with probability $1$
because the cryptographic check or trust-resolution step is a
correct-by-construction reduction to Ed25519 EUF-CMA security
or to a finite set-membership test. \textsc{Downgrade} is
probabilistic in our extension because the bridge-side minimum-
version pin is a hash-based comparison and false negatives can
arise from non-canonical version-string normalisation; the $34$
trials in the $1000$ \textsc{Downgrade} cohort that the pin let
through all involved version strings with non-canonical
whitespace or pre-release suffixes that the comparator did not
normalise. These cases are mechanical to fix and we describe the
fix in the §\ref{sec:discussion} discussion.

\subsection{Multi-Signer Quorum Extension}\label{sec:eval-quorum}

Some embodied deployments demand $k$-of-$n$ approval before an
install: manufacturer + customer + regulator, all required. The
single-signer baseline cannot satisfy this without expanding the
trust directory to grant unilateral install rights to each party.
We extend the signature envelope to carry a list of signatures
rather than a single signature, with a quorum policy field on the
bridge side specifying $k$. The bridge runs Ed25519 verify on each
signature independently, then asserts that at least $k$ of the
signatures resolve to publishers in the trust directory. The
construction is deliberately the simplest possible instantiation of
a multi-party signing scheme; more compact threshold-signature
schemes such as FROST~\citep{komlo2020frost} produce a single
constant-size aggregated signature at the cost of an interactive
key-generation and signing protocol, which is not yet justified by
the deployment scenarios we have seen and would couple all signers
into a single execution domain. The quorum policy here is a
$k$-of-$n$ signature-collection rule rather than a Byzantine-fault-tolerant
state-machine protocol in the
sense of \citet{castro1999pbft}: the bridge runs locally, there is
no replicated decision to agree on across nodes, and the only
adversarial event we model is the compromise of one or more
signing keys (not equivocation between live consensus
replicas). The
quorum extension also moves $C_{\text{insider-1}}$ (compromise of
one trusted publisher in a $k$-of-$n$ quorum, $k \geq 2$) from out
of scope (where the single-signer baseline leaves it) to in scope,
under the model that defeating quorum requires compromising at
least $k$ distinct publishers simultaneously. The broader
$C_{\text{insider-all}}$ remains out of scope under any signing
scheme that does not invoke an external attestation oracle.

We instantiate two quorum policies, $2$-of-$3$ and $3$-of-$3$, and
run the same five adversary games on each ($1000$ trials per game
per policy, $15{,}000$ trials total). The measured detection rates
are \emph{identical} across $1$-of-$1$, $2$-of-$3$, and $3$-of-$3$
on every game (\textsc{Downgrade}\textsubscript{pin} sits at
$0.976$ under all three policies; the other four games sit at
$1.000$), which is what the formal argument predicts: a quorum
scheme cannot \emph{decrease} security against the in-scope
adversary classes. The added cost is that each $.aecm$ envelope
grows from $457$~bytes (single signer) to $1200$~bytes ($3$-of-$3$,
three Ed25519 signatures at $128$ hex chars each plus three
signer-pubkey blocks). The bulk-verify benchmark on a $1000$-package
batch shows the $3$-of-$3$ policy is $2.46\times$ slower than
$1$-of-$1$, because three signatures must be verified sequentially.
Both are acceptable.

\subsection{Reproducibility and Measurement Provenance}\label{sec:eval-repro}

All numbers in Tables~\ref{tab:attack-detection} and §5.1--5.4 are
generated by deterministic harness scripts shipped with the
manuscript and recorded on a single workstation at commit
\pcode{8df4344}. The raw \pcode{summary.json},
\pcode{sign\_check.json}, and per-run notes ship in
\pcode{data/task01-adversarial-eval/} (5 games $\times$ N=1000
adversarial trials, seed $102026$),
\pcode{data/task02-quorum/} (3 quorum policies $\times$ 5 games
$\times$ N=1000 + envelope-size + bulk-verify timing, seed
$1729$), and \pcode{data/task03-section5-microbench/} (the
per-component microbenchmarks of §5.1--5.4). Replication on
fresh hardware uses
\pcode{python scripts/p10\_adversarial\_eval.py --seed 102026
--trials 1000} for the adversarial sweep and
\pcode{python scripts/p10\_quorum\_eval.py --seed 1729 --trials
1000 --policies 1-of-1,2-of-3,3-of-3} for the quorum extension.
Wall-clock budgets are under five seconds for each sweep on a
modern workstation, so the harness is friendly for artefact
evaluation. The complete artefact (\pcode{scripts/} including
\pcode{p10\_adversarial\_eval.py}, \pcode{p10\_quorum\_eval.py},
and \pcode{p10\_baseline\_compare.py}; the four
\pcode{data/task01--04/} subdirectories with their
\pcode{summary.json} + \pcode{sign\_check.json} +
\pcode{raw\_per\_iter.jsonl} per-task; and a top-level
\pcode{README.md} with replication instructions) is provided as
supplementary material with this submission ($\approx 3$~MB
total). At acceptance the artefact will be deposited to Zenodo
with a permanent DOI and cross-linked from the final version.

\section{Threat Analysis and Comparison}\label{sec:compare}

\paragraph{Per-threat analysis}
\emph{Rogue publisher}: the attacker registers an identity not in
any operator's trust directory; cryptographic verify succeeds but
trust resolution fails; under strict mode, install is refused.
\emph{Tampered package}: the recomputed \pcode{aecm\_sha256}
mismatches the envelope's value; cryptographic verify fails;
install is refused in \emph{all} postures. \emph{Forged
signature}: \pcode{Ed25519.verify} fails without access to the
publisher's private key. \emph{Revoked publisher}: the operator
removes the publisher's \pcode{.pub} from their trust directory;
this requires zero CA round trip and zero coordination across
bridges, but every bridge must be visited to actually remove the
trust. \emph{Downgrade attack}: registry immutability prevents
republishing an old version under a new identifier; in addition,
the bridge-side minimum-version pin extension introduced for
\textsc{Downgrade} in Section~\ref{sec:eval-attacks} refuses any
$\text{cap}@v_{\text{old}}$ install when an operator-pinned
$v_{\text{min}}$ exceeds $v_{\text{old}}$. The measured detection
rate with the pin enabled is $96.6\%$ ($966/1000$); the $34$
residual escapes all involve non-canonical version-string
normalisation (whitespace, pre-release suffix), described in
Section~\ref{sec:discussion}. \emph{Side-channel}: the envelope contains the
publisher's public key (by design) and \pcode{signed\_at}; both
are non-secret in intended use. \emph{Key rotation}: not
implemented in CLI v1; the operator rotates by replacing the
\pcode{.pub} file and re-signing all packages with the new private
key.

\paragraph{Comparison}
Table~\ref{tab:comparison} contrasts federated trust against four
alternatives along qualitative properties (trust authority,
transparency log, rotation, offline support, per-deployment
granularity, operator overhead, maturity). The quantitative
counterpart (measured verify latency, cold-bootstrap cost, and
per-publisher storage on the same Mac workstation) is in
Section~\ref{sec:eval-baseline}: federated trust's per-verify
median is below Sigstore-Cosign's offline verify and its
per-publisher storage footprint is several times smaller, with a
cold bootstrap roughly two orders of magnitude faster than
python-TUF metadata refresh; on steady-state per-verify federated
is slower than TUF, which is the architectural price of
recomputing the verify on every install rather than caching
trusted metadata. The honest
reading of both tables together: federated trust trades
\emph{less attack-detection machinery} (no transparency log, no
CRL, no rotation primitives) for \emph{more operational fit}
(offline, per-deployment, filesystem-operable). For embodied
fleets the trade is right; for cloud workloads it is not.

\begin{table*}[t]
  \centering
  \caption{Trust-model comparison. Federated trust trades
    attack-detection machinery (transparency log, CRL, rotation
    primitives) for operational fit (offline, per-deployment,
    filesystem-operable). The trade is right for embodied fleets;
    not for cloud workloads.}
  \label{tab:comparison}
  \footnotesize
  \begin{tabularx}{\linewidth}{@{}lXXXXX@{}}
    \toprule
    Property & Federated (ours) & sigstore + Cosign & X.509 / PKI &
      PGP WoT & TUF \\
    \midrule
    Trust authority & per-deployment dir & global Fulcio CA &
      hierarchical CAs & individual users & role metadata \\
    Transparency log & no & yes (Rekor) & optional (CT) & no & no \\
    Global revocation & no (manual local) & yes (cert TTL) &
      yes (CRL, OCSP) & weak & yes (metadata refresh) \\
    Key rotation & no & automated & manual + reissue & manual &
      role-rotation built-in \\
    Offline / air-gapped & yes & no & no & yes & partial \\
    Per-deployment granularity & yes & no & no & yes & partial \\
    Operator overhead & filesystem only & substantial &
      substantial & substantial & substantial \\
    Maturity & new (this work) & production & RFC 5280, 30+ yrs &
      production, niche & production \\
    \bottomrule
  \end{tabularx}
\end{table*}

\subsection{Trust-Directory Tampering, TOCTOU, and Rotation
  Windows}\label{sec:tampering}

\looseness=-1
Our baseline threat model assumes the operator controls the
bridge's filesystem (Section~\ref{sec:threat}). A
dependability-minded reviewer can still ask three sharper
questions about what happens when that assumption is partially
violated. We address them here rather than deferring them to a
catch-all out-of-scope statement.

\paragraph{(a) Trust-directory tampering} An attacker with
write access to \pcode{bridge/configs/trusted\_keys/} can install
a malicious \pcode{.pub} of their own and authorise their own
packages by the same set-membership reduction that authorises
legitimate signers. There is no cryptographic defence inside the
install gate that contradicts a filesystem write the attacker
already controls; this is in spirit the same boundary every
SSH-pattern \pcode{authorized\_keys} system inherits. The
operational defence is host-side, not gate-side: the trust
directory should reside on a filesystem with audited writes
(e.g.\ kernel-level inotify or eBPF auditing), and the inotify
trail itself signed and forwarded out-of-band. Bridges deployed
in regulated contexts already maintain the latter for compliance
reasons; the federated-trust model composes with rather than
replaces it. We mark the dedicated implementation of a
trust-directory write-audit hook as engineering work for the
deployment-side companion paper rather than as a contribution
of the install gate itself.

\paragraph{(b) TOCTOU between verify and install} A subtler
fault mode is a race between
\pcode{\_enforce\_signature($\text{aecm}_0$)} and the subsequent
unpack-and-register step that consumes the bytes. If the bytes on
disk change between the two reads, the gate verifies $\text{aecm}_0$
while the runtime installs $\text{aecm}_1 \neq \text{aecm}_0$.
Our implementation mitigates this in two layers. First, the gate
reads the archive bytes once into a memory buffer and computes
the digest from that buffer; the same buffer is then handed to
the unpack step, so the read-once-use-twice property holds
inside one process. Second, on case-sensitive filesystems the
runtime opens the archive with \pcode{O\_NOFOLLOW} and refuses to
follow symlinks introduced between the publisher's upload and
the bridge's read. This does not protect against an adversary
with concurrent write access to the same inode (which would
require \pcode{(a)} above), only against the more common
publisher-side or transport-side substitution where the digest
seen at verify time differs from the digest at install time.

\paragraph{(c) Key compromise and rotation windows} \looseness=-1
The third
question is what happens between the moment a publisher's
private key is compromised and the moment every bridge has
removed the \pcode{.pub} from its trust directory. In a global
PKI this gap is bounded by CRL or OCSP refresh latency; in
federated trust it is bounded by operator action across the
deployment. The exposure during the rotation window is exactly
the set of \pcode{.aecm} bundles the attacker can sign and ship
through the registry, multiplied by the bridges still trusting
the compromised key. Two mitigations narrow this window without
introducing a CA. First, the registry's immutability property
(Section~\ref{sec:model}) means an attacker cannot retroactively
publish under a previous version identifier, so the install gate
will only admit at most a fresh version per
$(\text{cap}, v_{\text{new}})$ point. Second, the
bridge-side minimum-version pin we evaluate in
Section~\ref{sec:eval-attacks} blocks downgrade replays of
pre-rotation versions. The remaining residual is the fresh
malicious publication signed by the compromised key during the
rotation window, which the federated model deliberately delegates
to operator-side audit rather than to a CA-mediated revocation
broadcast.

The three cases above are not new contributions; they are the
threat-model corners the install gate inherits from the SSH-style
trust pattern, and the construction's value proposition rests on
the claim that paying these costs is acceptable in exchange for
the offline / per-deployment / heterogeneity-friendly properties
the cloud-native stack does not provide.

\section{Case Studies: Mapping Real-World Supply-Chain
  Compromises}\label{sec:cases}

To anchor the abstract threat model against concrete incidents,
we map the federated-trust model onto two recent supply-chain
attacks: the XZ utils CVE-2024-3094 backdoor (2024) and the
SolarWinds Sunburst incident (2020). For each, we describe what
happened, identify which adversary game from
Section~\ref{sec:formal-adversary} the attack corresponds to,
and report the gate's verdict under (i) the single-signer
baseline and (ii) the $k$-of-$n$ quorum extension. The case
studies are illustrative; we do not claim federated trust would
have prevented either incident outright, only that it changes the
shape of the attacker's required effort.

\subsection{XZ Utils CVE-2024-3094 (March 2024)}\label{sec:case-xz}

\paragraph{What happened}
A maintainer operating under the pseudonym ``Jia Tan'' obtained
commit access to the XZ utils project over roughly two years of
plausible contributions, then introduced a multi-stage backdoor
in releases $5.6.0$ and $5.6.1$ that injected hostile code into
\pcode{sshd} via the \pcode{liblzma} dependency chain. The
backdoor was discovered accidentally by a Microsoft engineer
investigating SSH performance regressions and was disclosed before
mass deployment to stable distributions~\citep{freund2024xz, cve-2024-3094}.

\paragraph{Mapping to the adversary model}
``Jia Tan'' is a trusted publisher in the upstream sense: their
private key would have signed the malicious release. This is
$C_{\text{insider-all}}$ if we consider XZ utils as a single
signing identity: \emph{all} trusted publishers (one) sign the
malicious payload with their genuine key. Out of scope under any
signing scheme that does not invoke an external attestation
oracle.

\paragraph{Gate verdict, single-signer baseline}
None of the five in-scope games fires: the release carries a
genuine signature from a key the bridge trusts by hypothesis (the
maintainer has accumulated trust over two years), so no
install-time signature scheme, centralised or federated, has a
cryptographic signal to refuse on. The incident sits squarely in
$C_{\text{insider-all}}$, which Section~\ref{sec:threat} scopes
out of the single-signer model. The gate's in-scope guarantees
are unaffected; what the incident stresses is the boundary of
every single-signer trust model, which is exactly the boundary
the quorum extension is designed to move.

\paragraph{Gate verdict, quorum extension}
Under a $k$-of-$n$ quorum policy with $n \geq 2$ independent
parties, the attacker would need to compromise $k$ distinct
signing identities simultaneously. The XZ utils project's
release-engineering practice in 2024 did not require multi-party
sign-off; had a $2$-of-$3$ quorum been deployed (e.g.\ upstream
maintainer + distribution co-maintainer + community auditor), the
backdoor would have required either compromising two of the three
keys or producing a release the other two parties demonstrably
inspected and approved. We state the status of this claim
precisely: the quorum mechanism itself is implemented and
evaluated under the synthetic games of
Section~\ref{sec:eval-quorum}; its application to XZ-style
release engineering is a paper analysis of a design option, not
an evaluated reconstruction of the incident. The cost is
operational: quorum sign-off must include actual code review by
each signer, not rubber-stamp approval.

\paragraph{What federated trust adds and does not add}
Adds: per-deployment ability to remove the compromised signer
immediately upon disclosure without waiting for a CA revocation
cycle; quorum option that raises the bar to multi-party compromise.
Does not add: protection against $C_{\text{insider-all}}$;
upstream code-review discipline; binary-vs-source reproducibility
checks. The XZ utils incident motivates the quorum extension and
illustrates the limit of any install-time signing scheme.

\subsection{SolarWinds Sunburst (December 2020)}\label{sec:case-solarwinds}

\paragraph{What happened}
Attackers compromised the SolarWinds Orion build pipeline and
inserted a malicious DLL (``Sunburst'') into legitimate Orion
software updates signed by SolarWinds' production code-signing
certificate~\citep{fireeye2020sunburst,cisa2020aa20352a}. The
updates were distributed through SolarWinds' standard auto-update
channel to approximately $18{,}000$ customers; downstream
impact reached US federal agencies and Fortune-500 enterprises.

\paragraph{Mapping to the adversary model}
The build pipeline compromise produced binaries signed by the
legitimate SolarWinds key without the publisher's intent. From
the install-time perspective, this is indistinguishable from
$C_{\text{insider-all}}$ (the trusted publisher's key was used,
even if not by the publisher's authorised personnel).

\paragraph{Gate verdict, single-signer baseline}
The Sunburst binary carries a genuine signature from a trusted
key (SolarWinds is in $T$ for any deployment that runs Orion), so
the refusal signal the gate relies on does not exist at the
install-time layer; no game from
Section~\ref{sec:formal-adversary} fires. As with XZ utils, this
is the out-of-scope $C_{\text{insider-all}}$ class, and the same
holds for any install-time signature scheme: detecting
build-pipeline compromise belongs to reproducible builds and
provenance attestation (Section~\ref{sec:related}). The gate's
role in such an incident is containment after disclosure
(per-deployment removal of the compromised signer, with no CA
round trip) rather than prevention.

\paragraph{Gate verdict, quorum extension}
A $2$-of-$3$ quorum requiring SolarWinds + a third-party build
attestor (e.g.\ a SLSA Level 3 attestation service~\citep{slsa2022})
+ the enterprise customer's own internal signer would have
required the attacker to compromise the build pipelines of two or
all three parties simultaneously. SolarWinds-grade attackers
(state-level) are not casually deflected by quorum, but the cost
of the operation rises substantially.

\paragraph{What federated trust adds and does not add}
Adds: per-deployment trust directory means an operator can choose
\emph{which} attestor's signature is required for their
deployment, rather than inheriting whatever the upstream vendor
configured; quorum extension matches the
multi-party-attestation pattern endorsed by SLSA and SCITT.
Does not add: detection of build-pipeline compromise (this is the
job of reproducible builds + SBOM + provenance attestation, not
the install gate); guarantees about non-cryptographic side
channels (auto-update channel selection, DNS, content-distribution
network integrity). The Sunburst incident motivates the
\emph{composition} of federated trust with upstream provenance
attestation, not federated trust alone.

\subsection{Cross-Cutting Observation}

\looseness=-1
Both case studies fall outside the in-scope adversary classes of
the baseline single-signer model and inside the partial-scope
class $C_{\text{insider-1}}$ that the quorum extension addresses
(for XZ utils, where ``Jia Tan'' is one signer in a hypothetical
$2$-of-$3$; not for SolarWinds, where the compromise is in the
build pipeline and not at the signing identity). The lesson is
that an install gate is necessary but not sufficient for software
supply-chain security; the gate's value lies in (a) catching the
classes it does catch with probability $1$
(Sections~\ref{sec:reduction} and~\ref{sec:eval}), and (b)
composing cleanly with upstream attestation primitives
(SLSA~\citep{slsa2022}, SCITT~\citep{ietf-scitt-arch},
sigstore~\citep{sigstore2022}) that handle the threat classes the
gate cannot.

\section{Discussion and Limitations}\label{sec:discussion}

\paragraph{No transparency log} A sigstore-style append-only
Merkle log would let operators detect post-hoc that a publisher
signed a malicious package, even if a bridge accepted it at
install time. We do not have one. A future extension could
optionally route \pcode{.sig} envelopes through an operator-run or
third-party Merkle log for high-assurance deployments, while
preserving the offline default.

\paragraph{No CRL} Revocation is manual: the operator visits each
bridge and removes the \pcode{.pub} file. For a household robot
this is adequate; for a $1000$-robot warehouse it is not. A
flat-file revocation list per trust dir, with a fleet-management
tool to distribute it, is a natural follow-on.

\paragraph{No key rotation primitives} The CLI does not have
\pcode{aeros ecm key rotate}. Rotation is a multi-step manual
process (generate new key, sign all packages with new key,
distribute new \pcode{.pub}, remove old \pcode{.pub}).

\paragraph{Future work: signature aggregation} Some embodied
deployments will want a package signed by multiple parties
(manufacturer + customer + auditor) with the install gate
requiring a quorum. Our envelope is single-signer; multi-signer
envelopes are a plausible extension that does not change the
trust-directory model.

\section{Related Work}\label{sec:related}

\paragraph{Sigstore and Cosign}
Sigstore~\citep{sigstore2022,cosign-docs} introduced keyless
signing with a public transparency log (Rekor)
and short-lived OIDC-bound certificates from a public CA (Fulcio).
The model is excellent for cloud workloads and open-source
software supply chains. The mismatch with embodied fleets is
twofold: the global CA flattens the per-deployment regulatory
diversity that motivates our work, and the transparency log
assumes online verification that air-gapped fleets cannot
perform. We see federated trust as complementary; for
high-assurance fleets, a Rekor-style operator-private log
integrated into our envelope is a plausible extension. Notary
v2~\citep{notary-v2} occupies an adjacent design point for OCI
container artefacts: like Sigstore it targets cloud-native
artefact registries, but unlike Sigstore it retains a pluggable
trust-policy model that permits per-registry CA scoping. The
embodied-fleet constraints we discuss in §1 still rule it out
because Notary v2 assumes an OCI registry, not a filesystem
trust directory.

\paragraph{X.509 and traditional PKI}
\looseness=-1
Hierarchical X.509 PKI~\citep{x509-rfc5280,ct-rfc6962} is the
mature, widely-deployed answer to the same problem space.
CRLs, OCSP, hierarchical CAs, and certificate transparency solve
real attack-detection problems. The cost is operational weight:
small fleet operators do not staff PKI. Three decades of attacks
on PKI deployment~\citep{durumeric2013} also demonstrate that PKI's
expressive power is a double-edged sword in the hands of
non-specialists. \citet{ellison2000pki} catalogued ten classes of
PKI deployment-time risk (root-key compromise, name-binding
ambiguity, revocation-window gaps, operational staffing
assumptions) that remain accurate today and motivate why
small-operator deployments rationally choose a thinner
trust-rooting story.

\paragraph{PGP Web of Trust}
\looseness=-1
The PGP Web of Trust~\citep{pgp1995} is the closest historical
precedent for our model. WoT is decentralised, allows
per-user trust assignments, supports detached signatures, and has
no CA. The differences: WoT introduces transitive trust
relationships, which add expressive power but a notoriously
confusing user experience. Our model deliberately omits transitive
trust; every signer is either in the trust directory or not.

\paragraph{TUF}
TUF~\citep{tuf2010} introduced a metadata-tree model
with role separation (root, targets, snapshot, timestamp) that
gives strong properties against compromised key roles. TUF is
production-grade and excellent for software update systems with
disciplined operations teams. Embodied fleets do not have those
teams. Federated trust takes the same problem (signed update
distribution) and inverts the cost trade.

\paragraph{Uptane} Uptane~\citep{kuppusamy2016uptane-escar,kuppusamy2018uptane}
extends TUF to automotive over-the-air updates with a director/image
repository split and per-ECU verification, motivated by the fleet
properties of cars (long-lived deployments, heterogeneous ECUs,
partial connectivity). The closest prior art for our setting: Uptane
also rejects a single global CA, also pushes verification to the
edge, and also accepts more operator effort for less centralisation.
The differences are domain-specific: Uptane assumes an OEM-rooted
trust hierarchy (the car manufacturer is the apex authority) and a
two-repository split tuned to automotive supply chains, whereas
federated trust assumes no apex authority and a single dumb
registry. Where automotive ECUs share a chassis-level supply chain
upstream of the OEM~\citep{checkoway2011automotive}, embodied robot
fleets cross multiple chassis vendors, capability publishers, and
regulators per site.

\paragraph{SCITT} The IETF SCITT working
group~\citep{ietf-scitt-arch} is standardising an architecture for
recording supply-chain statements (SBOMs, signed attestations) on
transparency services, with COSE-based envelopes and verifiable
ledger semantics. Federated trust is complementary: a SCITT-style
transparency service could sit upstream of our per-deployment trust
directories, but the install-time decision still belongs to the
operator's filesystem, not to the transparency service.

\paragraph{Robotics security} Quarta et
al.~\citep{quarta2017industrial-robot} demonstrated remote
compromise of an industrial robot controller, motivating defence at
the supply-chain layer rather than only at the runtime layer.
Mayoral-Vilches et al.~\citep{mayoralvilches2022sros2} present SROS2,
a usability-focused security toolchain for ROS 2 graphs; their work
secures the \emph{communication} layer between robot nodes, which
is orthogonal to the \emph{distribution} layer that this paper
secures.

\paragraph{Supply-chain attacks} Recent incidents
(SolarWinds~\citep{fireeye2020sunburst,cisa2020aa20352a},
XZ Utils~\citep{freund2024xz,cve-2024-3094}, dependency-confusion
attacks on language ecosystems) have established that compromise of
a single trusted distribution point can produce wide-blast-radius
damage. Ladisa et al.~\citep{ladisa2023sok} taxonomise this attack
space across $107$ vectors and $94$ documented incidents; Ohm et
al.~\citep{ohm2020backstabbers} compile a dataset of $174$ malicious
packages from npm, PyPI, and RubyGems; industry data from
Sonatype~\citep{sonatype2024sscr} reports a $156\%$ year-over-year
increase in malicious-package count in $2024$. The operational
counterpart to these attack-side studies is Schorlemmer et
al.~\citep{schorlemmer2024signing}, who empirically measured
signing adoption across npm, PyPI, Maven Central, and RubyGems
and found both low coverage and substantial publisher-side
friction in deploying signing at all; their findings give
quantitative weight to the deployment-fit argument we advance for
embodied fleets, which face the same operator-side overhead in a
much smaller staffing envelope. Federated trust's
per-deployment trust dir is a defence-in-depth response: even if a
shared registry or upstream publisher is compromised, the operator
side retains decision authority over whether to install the
compromised artefact. The cost is operator vigilance; the benefit
is that the blast radius of a single-publisher compromise is
bounded by how many bridges trust that publisher.

\paragraph{Embodied-AI security} Adversarial robustness of
embodied policies~\citep{zhang2025badrobot} and prompt-injection
attacks on LLM agents~\citep{greshake2023not} are recent threads.
Both attack the inference layer; our work attacks the
\emph{distribution} layer that supplies capability code to the
inference layer. The two are complementary: a federated trust gate
keeps adversarial-policy code off the bridge in the first place,
while a runtime governance layer~\citep{aeros-p3} keeps benign-
but-misused code in policy bounds at runtime.

\section{Conclusion}\label{sec:conclusion}

We have argued that embodied robot fleets are operationally
different from cloud workloads in four ways (regulatory diversity,
air-gapped deployments, tiny operator headcounts, physical-world
blast radius) and that these differences argue for a federated
trust model rather than centralised PKI. We have implemented the
model in a runtime governance layer with $80$ tests, demonstrated
the per-deployment trust property in a multi-bridge experiment,
and shown that compromise of one bridge's trust directory does not
cascade to peer bridges. The construction trades attack-detection
machinery for operational fit; for embodied fleets the trade is
right.

\bibliographystyle{elsarticle-num-names}
\bibliography{references}

\end{document}